\documentclass[preprint,aps,amsmath,superscriptaddress,nofootinbib,tightenlines,showpacs]{revtex4}
\usepackage{txfonts}
\usepackage{mathrsfs}
\usepackage{amsmath}
\usepackage{amsfonts}
\usepackage{amssymb}
\usepackage{latexsym}
\usepackage{graphicx}
\usepackage{bm}
\usepackage{epsfig}
\usepackage[english]{babel}
\def \be {\begin{equation}}
\def \ee {\end{equation}}
\def \bea {\begin{eqnarray}}
\def \eea {\end{eqnarray}}
\def \nn {\nonumber}

\def \a {\alpha}
\def \b {\beta}

\def \d {\delta}

\def \m {\mu}
\def \n {\nu}
\def \k {\kappa}

\def \s {\sigma}
\def \r {\rho}
\def \o {\omega}

\def \th {\theta}
\def \Th {\Theta}

\def \t {\tau}
\def \dag {\dagger}
\def \p {\partial}

\def\bd{\begin{document}}
\def\ed{\end{document}}
\def\nn{\nonumber}
\def\bea{\begin{eqnarray}}
\def\eea{\end{eqnarray}}
\let\bm=\bibitem
\let\la=\label

\def\N{{\cal N}}
\def\sst{\scriptscriptstyle}
\def\thetabar{\bar\theta}
\def\Tr{{\rm Tr}}
\def\one{\mbox{1 \kern-.59em {\rm l}}}

\def\a{\alpha}      \def\da{{\dot\alpha}}
\def\b{\beta}       \def\db{{\dot\beta}}
\def\c{\gamma}  \def\C{\Gamma}  \def\cdt{\dot\gamma}
\def\d{\delta}  \def\D{\Delta}  \def\ddt{\dot\delta}
\def\e{\epsilon}        \def\vare{\varepsilon}
\def\f{\phi}    \def\F{\Phi}    \def\vvf{\f}
\def\h{\eta}
\def\k{\kappa}
\def\l{\lambda} \def\L{\Lambda}
\def\m{\mu} \def\n{\nu}
\def\o{\omega}
\def\P{\Pi}
\def\r{\rho}
\def\s{\sigma}  \def\S{\Sigma}
\def\t{\tau}
\def\th{\theta} \def\Th{\Theta} \def\vth{\vartheta}
\def\X{\Xeta}
\def\z{\zeta}
\def\w{\wedge}
\def\u{\underline}
\def\hs{\hspace}

\def\cA{{\cal A}} \def\cB{{\cal B}} \def\cC{{\cal C}}
\def\cD{{\cal D}} \def\cE{{\cal E}} \def\cF{{\cal F}}
\def\cG{{\cal G}} \def\cH{{\cal H}} \def\cI{{\cal I}}
\def\cJ{{\cal J}} \def\cK{{\cal K}} \def\cL{{\cal L}}
\def\cM{{\cal M}} \def\cN{{\cal N}} \def\cO{{\cal O}}
\def\cP{{\cal P}} \def\cQ{{\cal Q}} \def\cR{{\cal R}}
\def\cS{{\cal S}} \def\cT{{\cal T}} \def\cU{{\cal U}}
\def\cV{{\cal V}} \def\cW{{\cal W}} \def\cX{{\cal X}}
\def\cY{{\cal Y}} \def\cZ{{\cal Z}}

\def\ua{\underline{\alpha}} \def\ubb{\underline{\beta}}
\def\ug{\underline{\gamma}}
\def\ub{\underline{\phantom{\alpha}}\!\!\!\beta}
\def\uc{\underline{\phantom{\alpha}}\!\!\!\gamma}
\def\um{\underline{\mu}} \def\un{\underline{\nu}}
\def\ud{\underline\delta}
\def\ue{\underline\epsilon}
\def\una{\underline a}\def\unA{\underline A}
\def\unb{\underline b}\def\unB{\underline B}
\def\unc{\underline c}\def\unC{\underline C}
\def\und{\underline d}\def\unD{\underline D}
\def\une{\underline e}\def\unE{\underline E}
\def\unf{\underline{\phantom{e}}\!\!\!\! f}\def\unF{\underline F}
\def\unm{\underline m}\def\unM{\underline M}
\def\unn{\underline n}\def\unN{\underline N}
\def\unp{\underline{\phantom{a}}\!\!\! p}\def\unP{\underline P}
\def\unq{\underline{\phantom{a}}\!\!\! q}
\def\unQ{\underline{\phantom{A}}\!\!\!\! Q}
\def\unH{\underline{H}}
\def\ul{\underline}

\def\As {{A \hspace{-6.4pt} \slash}\;}
\def\bs {{b \hspace{-6.4pt} \slash}\;}
\def\Ds {{D \hspace{-6.4pt} \slash}\;}
\def\ds {{\del \hspace{-6.4pt} \slash}\;}
\def\ss {{\s \hspace{-6.4pt} \slash}\;}
\def\ks {{ k \hspace{-6.4pt} \slash}\;}
\def\ps {{p \hspace{-6.4pt} \slash}\;}
\def\pas {{{p_1} \hspace{-6.4pt} \slash}\;}
\def\pbs {{{p_2} \hspace{-6.4pt} \slash}\;}

\def\Fh{\hat{F}}
\def\Vh{\hat{V}}
\def\Xh{\hat{X}}
\def\ah{\hat{a}}
\def\xh{\hat{x}}
\def\yh{\hat{y}}
\def\ph{\hat{p}}
\def\xih{\hat{\xi}}

\def\psit{\tilde{\psi}}
\def\Psit{\tilde{\Psi}}
\def\tht{\tilde{\th}}

\def\At{\tilde{A}}
\def\Qt{\tilde{Q}}
\def\Rt{\tilde{R}}
\def\Nt{\tilde{N}}

\def\at{\tilde{a}}
\def\st{\tilde{s}}
\def\ft{\tilde{f}}
\def\pt{\tilde{p}}
\def\qt{\tilde{q}}
\def\vt{\tilde{v}}
\def\nt{\tilde{n}}

\def\delb{\bar{\partial}}
\def\bz{\bar{z}}
\def\bD{\bar{D}}
\def\bB{\bar{B}}

\def\bk{{\bf k}}
\def\bl{{\bf l}}
\def\bp{{\bf p}}
\def\bq{{\bf q}}
\def\br{{\bf r}}
\def\bx{{\bf x}}
\def\by{{\bf y}}
\def\bR{{\bf R}}
\def\bV{{\bf V}}

\def\d{\delta}\def\D{\Delta}\def\ddt{\dot\delta}

\def\p{\partial} \def\del{\partial}
\def\xx{\times}
\def\uno{\mbox{1 \kern-.59em {\rm l}}}

\def\trp{^{\top}}
\def\inv{^{-1}}
\def\dag{{^{\dagger}}}
\def\pr{\prime}

\def\rar{\rightarrow}
\def\lar{\leftarrow}
\def\lrar{\leftrightarrow}

\def\cw{{\cal W}}
\def\cz{{\cal Z}}
\def\tcm{\tilde{\cal M}}
\def\sgn{{\rm sgn}}
\def\sd {d^{4|4}}
\def\lan{\langle}
\def\ran{\rangle}

\def\tr{\mbox{tr}}
\def\sign{\mbox{sign}}
\def\fnl{f_\text{NL}}
\def\horava{Ho\v{r}ava}
\def\la{\langle}
\def\ra{\rangle}
\def\mb{\mathbf}
\def\nn{\nonumber}
\def\hl{Ho\v{r}ava-Lifshitz}
\def\p{\partial}
\def\dij{\delta_{ij}}
\def\tr{\mbox{tr}}
\def\sign{\mbox{sign}}
\def\fnl{f_\text{NL}}
\def\horava{Ho\v{r}ava}
\def\la{\langle}
\def\ra{\rangle}
\def\mb{\mathbf}
\def\nn{\nonumber}
\def\hl{Ho\v{r}ava-Lifshitz}
\def\p{\partial}
\def\dij{\delta_{ij}}

\begin{document}


\title{Static Spherically Symmetric Solutions to modified \horava-Lifshitz Gravity with Projectability Condition }
\author{Jin-Zhang Tang\footnote{Electronic address:JinzhangTang@pku.edu.cn},\hspace{2ex} Bin Chen\footnote{Electronic
address: bchen01@pku.edu.cn}} \affiliation{Department of Physics,
and State Key Laboratory of Nuclear Physics and Technology, Peking
University, Beijing 100871, China}

\date{\today\\ \vspace{1cm}}
\begin{abstract}
In this paper we seek static spherically symmetric solutions of
\horava-Lifshitz-like gravity with projectability condition. We
consider the most general form of gravity action without detailed
balance, and require the spacetime metric to respect the
projectability condition. We find that for any value of $\l$, it may
exists the solutions of topology $\mathbb{R}\times \mathbb{M}_3$,
where $\mathbb{R}$ is the time direction and $\mathbb{M}_3$ is a
three-dimensional maximally symmetric space depending on the value
of cosmological constant and the potential of the action. Besides,
in the UV region where $\lambda\neq 1$, we find Minkowski or
de-Sitter space-time as the solution, while in the IR region where
$\lambda=1$, we prove that (dS-)Schwarzschild solution is the only
nontrivial solution.
 We also notice that the other static spherically symmetric solutions
found in the literature do not satisfy the projectability condition
and are not the solutions we get. Our study shows that in
\horava-Lifshitz gravity with projectability condition, there is no
novel correction to Einstein's general relativity  in solar system
tests.

\pacs{98.80.Cq}
\end{abstract}
\maketitle

\section{introduction}\label{sec-intro}

Diffeomorphism is an essential symmetry of Einstein's relativity
theory of gravity. It has been widely believed to be exact in any
theory of gravity. However, in the recent proposal by
\horava\cite{Horava:2008ih,Horava:2009uw} on gravity theory, it is
no longer an exact symmetry. The basic idea behind \horava's
theory is that time and space may have different dynamical scaling
in UV limit. This was inspired by the development in quantum
critical phenomena in condensed matter physics, with the typical
model being Lifshitz scalar field
theory\cite{Lifshitz,Chen:2009ka}. In this \horava-Lifshitz
theory, time and space will take different scaling behavior as
\begin{equation}\label{scaling}
    \mb{x}\rightarrow b\mb{x},\;\;\;\; t\rightarrow b^zt,
\end{equation}
where $z$ is the dynamical critical exponent characterizing the
anisotropy between space and time. Due to the anisotropy, instead of
diffeomorphism, we have the so-called foliation-preserving
diffeomorphism. The transformation is now just
 \bea
 t&\rightarrow& \tilde{t}(t), \nn\\
 x^i &\rightarrow& \tilde{x^i}(x^j,t).
 \eea
As a result, there is one more dynamical degree of freedom in
\hl-like gravity than in the usual general relativity. Such a degree
of freedom could play important role in UV physics, especially in
early cosmology\cite{Cai:2009dx,Chen:2009jr}. At IR, due to the
emergence of new gauge symmetry, this degree of freedom is not
dynamical any more such that the kinetic part of the theory recovers
the one of the general relativity.

Since time direction plays a privileged role in the whole
construction, it is more convenient to work with ADM metric
\begin{equation}\label{ADMmetric}
    ds^2=-N^2dt^2+g_{ij}(dx^i+N^idt)(dx^j+N^jdt),
\end{equation}
in which $N$ and $N_i$ are called ``lapse" and ``shift" variables
respectively. Then we have the following transformations on the
metric components:
 \bea
 \delta g_{ij}&=&\p_i\xi^kg_{jk}+\p_j\xi^k
 g_{ik}+\xi^k\p_kg_{ij}+\xi^0 \dot{g}_{ij} \nn\\
 \delta N_i&=&\p_i\xi^jN_j+\xi^j\p_j
 N_i+\dot{\xi}^jg_{ij}+\dot{\xi}^0N_i+\xi^0\dot{N}_i \nn\\
 \delta N&=&\xi^j\p_j N +\dot{\xi}^0N+\xi^0\dot{N} \label{gaugetr}
 \eea
It seems natural to choose the lapse function $N$ to be projectable
function on the spacetime foliation, i.e. only a function of $t$.
Such a choice makes the above gauge transformations simpler and more
transparent. More importantly, with the projectable condition, in
the
 Hamiltonian formulation the constraints could form a closed
 algebra \cite{Horava:2008ih} since the momentum conjugate to $N$
 does not lead to a local constraint. On the contrary, if the
 projectable condition on $N$ is abandoned, then the theory would
 not be
 well-defined, as shown in \cite{Horava:2008ih,Li:2009bg}.
 Therefore in this letter, we will focus on the case with
the projectable condition.

Taken \hl~ gravity as a new gravitational theory, it is an
important issue to study its static spherically symmetric
solutions. This issue has been widely studied in the literature,
see
\cite{Lu2009,Nastase2009,Kehagias:2009is,AhmadGhodsi2009,Colgain:2009fe,park2009}.
In these papers, for example \cite{Lu2009,park2009}, it was
assumed that the metric of the black solutions had the following
form
\begin{equation}\label{Nr}
ds^2=-N(r)^{2}dt_{S}^2+\frac{dr^2}{g(r)}+r^{2}(d\theta^{2}+\sin^{2}\theta
d\phi^2).
\end{equation}
From this metric ansatz, it was found that there were new
spherically symmetric solutions, even at IR. For example, in
\cite{Kehagias:2009is}, based on a modified \hl~-type action, an
asymptotically flat solution with
 \be
 g=N^2=1+\omega r^2-\sqrt{r(\o^2r^3+4\o M)}
 \ee
 was found.
 This raised the issue that if
there is any observational effect in solar system tests\cite{Solar}.

However, in the above ansatz (\ref{Nr})  the ``lapse function''
$N(r)$ obviously breaks the ``projectability condition''. As the
\horava\ gravity is only well defined when the ``projectability
condition'' is preserved, this naturally leads one to ask whether
the above new solutions still are the solutions of \hl~gravity
with the projectability condition after proper coordinates
transformation? The answer to this  question is not obvious,
considering the freedom in doing coordinate transformation. For
instance, a static
 spherically symmetric solution in the flat spacetime  could be represented
 in Schwarzschild coordinates as
\begin{equation}\label{SchWarzs01}
ds^{2}=-(1-\frac{2GM}{r})dt_{S}^{2}+(1-\frac{2GM}{r})^{-1}dr^{2}+r^{2}\left(d\theta^{2}+\sin^{2}\theta
d\phi^{2}\right),
\end{equation}
which looks against the projectability condition. By a
transformation into the Painlev\'e-Gullstrand
 coordinates\cite{Painleve,Gullstrand,Lematre,Hawking}
\begin{equation}
dt_{S}=dt_{PG}\mp \frac{\sqrt{2GM/r}}{1-2GM/r}dr,
\end{equation}
the solution \eqref{SchWarzs01} becomes
\begin{equation}
ds^{2}=-dt_{PG}^{2}+(dr\pm\sqrt{\frac{2GM}{r}}dt_{PG})^{2}+r^{2}\left(d\theta^{2}+\sin^{2}\theta
d\phi^{2}\right).
\end{equation}
Comparing with the ADM metric \eqref{ADMmetric}, we find that the
``lapse function'' $N=1$, which is in accord with the
``projectability condition''.

Furthermore, we would like to know if there are any other new
solutions, especially at IR, which may have significant physical
implication in IR physics. Therefore, in this letter, we study the
static spherically symmetric solutions to modified
\horava-Lifshitz gravity with the projectability condition. We
consider the most general form of the action without the detailed
balance condition. We find that for any value of $\l$, if the
potential term is properly chosen, there may exists the solutions
of topology $\mathbb{R}\times \mathbb{M}_3$, where $\mathbb{R}$ is
the time direction and $\mathbb{M}_3$ is a three-dimensional
maximally symmetric space. In the case without the cosmological
constant in the action, $\mathbb{M}_3$ is just the flat spacetime.
In the case with the cosmological constant, $\mathbb{M}_3$ could
be a three-dimensional sphere $\mathbb{S}^3$ or hyperboloid
$\mathbb{H}^3$, depending on the potential. Moreover, apart from
these solutions, in the UV region where $\lambda\neq1$, we find
either de-Sitter space-time or Minkowski spacetime, up to the
cosmological constant, while in the IR region where $\lambda=1$,
we prove that (dS)-Schwarzschild solution is the only nontrivial
solution. This result seems in accordence with
\cite{A.A.Kocharyan}.
 We also notice that the other static spherically symmetric solutions
found in the literature do not satisfy the projectability condition
and are not the solutions we want. Our study shows that in
\horava-Lifshitz-like Gravity with the projectability condition,
there is no novel correction to Einstein's general relativity  in
solar system tests.

We study the topological static spherically symmetric solutions in
the \hl-like gravity as well. We choose the metric ansatz in which
$d\Omega_{k}^{2}$ denotes the line element for an 2-dimensional
Einstein space with constant scalar curvature $2k$. Without loss
of generality, one may take $k=0,\pm1$ respectively. The $k=1$
case has been discussed above. To $k=-1$ case, we find that it may
also exists the solutions of topology $\mathbb{R}\times
\mathbb{M}_3$ for all $\lambda$. In the UV region where
$\lambda\neq 1$, the only possible solution is either Minkowski or
de-Sitter space-time with topological twist. In the IR region
where $\lambda=1$, the Schwarzschild topological black hole is the
only nontrivial solution. For the case $k=0$, there is not a
Schwarzschild solution at IR or de-sitter space-time in the UV
region because $f$ can't be zero.

\section{The modified \hl~ gravity}\label{sec-modification}

In this section, we give a brief review of \hl~gravity and its
modifications. Using the ADM formalism, the action of this
\horava-Lifshitz gravitational theory is given
by\cite{Horava:2008ih,Horava:2009uw}

\begin{eqnarray}\label{S_g:origin}
\nn
S&=&\int dtd^{3}\mb{x}(\mathcal{L}_{K}+\mathcal{L}_{V}),\\
\nn
\mathcal{L}_{K}&=&\sqrt{g}N\left\{\frac{2}{\kappa^2}(K_{ij}K^{ij}-\lambda
K^2)\right\},\\
\nn
\mathcal{L}_{V}&=&\sqrt{g}N\left\{\frac{\kappa^2\mu^2(\Lambda_{W}R-3\Lambda^2_{W})}{8(1-3\lambda)}+\frac{\kappa^2\mu^2(1-4\lambda)}{32(1-3\lambda)}R^2\right.\\
&&\left.-\frac{\kappa^2}{2\omega^4}\left(C_{ij}-\frac{\mu\omega^2}{2}R_{ij}\right)\left(C^{ij}-\frac{\mu\omega^2}{2}R^{ij}\right)\right\},
\end{eqnarray}
where $\mathcal{L}_{K}$ is the kinetic term and $\mathcal{L}_{V}$ is
the potential term. In the action, $\lambda,\kappa,\mu,\omega$ and
$\Lambda_{W}$ are the coupling parameters, and $C_{ij}$ is the
Cotton tensor defined by
\begin{equation}
C^{ij}=\epsilon^{ikl}\nabla_{k}\left(R^j_l-\frac{1}{4}R\delta^j_l\right).
\end{equation}

 The study of the perturbations around
the Minkowski vacuum shows that there is ghost excitation when
$\frac{1}{3}<\lambda<1$. This indicates that the theory is only
well-defined in the region $\l\leq \frac{1}{3}$ and $\l\geq 1$.
Since the theory should be RG flow to IR with $\l =1$, we expect
that at UV, $\l >1$ to have a well-defined RG flow. At IR, $\l =1$,
the kinetic term recovers the one of standard general relativity.
Comparing to the action of the general relativity in the ADM
formalism, the speed of light, the Newton's constant and the
cosmological constant emerge as
\begin{eqnarray}\label{c:former}
  c = \frac{\kappa^2\mu}{4}\sqrt{\frac{\Lambda_W}{1-3\lambda}},
  \hspace{3ex}
  G = \frac{\kappa^2}{32\pi c},
  \hspace{3ex}\Lambda=\frac{3}{2}\Lambda_W.
\end{eqnarray}
It follows from \eqref{c:former} that for $\lambda>1/3$ ,the
cosmological constant $\Lambda_{W}$ has to be negative. It was
noticed in \cite{Lu2009} that  if we make an analytic continuation
of the parameters
\begin{equation}
\mu \to i\mu, \hspace{4ex} \omega^2 \to -i\omega^2,
\end{equation}
the four-dimensional action remains real. In this case, the emergent
speed of light becomes
\begin{equation}
c= \frac{\kappa^2\mu}{4}\sqrt{\frac{\Lambda_W}{3\lambda-1}}.
\end{equation}
The requirement that this speed be real implies that $\Lambda_{W}$
must be positive for $\lambda>\frac{1}{3}$.

One important feature of original \hl\  gravity is that it
respects the so-called ``detailed balance''
condition\cite{Horava:2008ih,Horava:2009uw}. However, it turns out
that the detailed balance condition is not essential to the
theory. It could be just a nice way to organize the action.  If
abandoning `detailed balance'' and just requiring the model to be
power-counting renormalizable, we find that the most general form
of the action is of the form \cite{Visser_2009}
\begin{eqnarray}\label{S_g:General_model01}
\nn
S&=&\int dtd^{3}\mb{x}(\mathcal{L}_{K}+\mathcal{L}_{V}),\\
\nn
\mathcal{L}_{K}&=&\sqrt{g}N\left\{g_{K}(K_{ij}K^{ij}-\lambda K^2)\right\},\\
\nn
\mathcal{L}_{V}&=&\sqrt{g}N\left\{-g_{0}\zeta^{6}+g_{1}\zeta^{4}R+g_{2}\zeta^{2}R^{2}+g_{3}\zeta^{2}R_{ij}R^{ij}\right.\\
\nn
&&\left.+g_{4}R^{3}+g_{5}R(R_{ij}R^{ij})+g_{6}R^{i}_{j}R^{j}_{k}R^{k}_{i}\right.\\
&&\left.+g_{7}R\nabla^{2}R+g_{8}\nabla_{i}R_{jk}\nabla^{i}R^{jk}\right\}.
\end{eqnarray}
where $\zeta$ is a suitable factor to ensure the couplings $g_{a}$
are all dimensionless. From anisotropic scaling counting, five of
these operators are marginal(renormalizable) and four are
relevant(super-renormalizable). And we can rescale the time and
space coordinates to set both $g_{K}\to 1$ and $g_{1}\to1$ without
loss of generality. In the following, we will study the static
spherically symmetric solution to the action
\eqref{S_g:General_model01}.

\section{Static spherically symmetric solutions}

The static spherically symmetric solutions of \horava-Lifshitz
gravity have been discussed by
\cite{Lu2009,Nastase2009,Kehagias:2009is,AhmadGhodsi2009,park2009}.
In these paper, it was assumed that the metric of the solutions took
the form \eqref{Nr}. Consequently, some new kinds of solutions have
been found. For the Horava's original model,  three types of
solutions were found in \cite{Lu2009}. The first one is given by
\begin{equation}\label{Lu'sSolution01}
g=1+x^{2},\;\;\; x=\sqrt{-\Lambda_{W}}r,
\end{equation}
without any restriction on the function $N(r)$. This is valid for
all $\lambda$. And the other two solutions are given by
\begin{equation}\label{Lu'sSolution02}
g=1+x^{2}-\alpha
x^{\frac{2\lambda\pm\sqrt{6\lambda-2}}{\lambda-1}},\;\;\;\;N=x^{-\frac{1+3\lambda\pm2\sqrt{6\lambda-2}}{\lambda-1}}g,
\end{equation}
where $\alpha$ is an integration constant. For the solution to be
real, it is necessary that $\lambda>1/3$.

In paper \cite{park2009}, Park got a more general solution in the IR
region when $\lambda=1$, basing on an action softly breaking the
detailed balance condition
\begin{equation}\label{Park's solution}
N^{2}=g=1+(\omega-\Lambda_{W})r^{2}-\sqrt{r\left[\omega\left(\omega-2\Lambda_{W}\right)r^{3}+\beta\right]}.
\end{equation}
Certainly, for a general form of the action like
\eqref{S_g:General_model01}, it may exists other kinds of solution
with the metric ansatz \eqref{Nr}.

For the metric of the form \eqref{Nr}, we can work in the
Painlev\'e-Gullstrand coordinates by making a transformation
\begin{equation}
dt_{S}=dt_{PG}-\frac{\sqrt{1-N^{2}}}{N^{2}}dr.
\end{equation}
 Then the ansatz \eqref{Nr} becomes
\begin{equation}
ds^{2}=-dt_{PG}^{2}+(dr+\sqrt{1-N^{2}}dt_{PG})^{2}+(\frac{1}{g}-\frac{1}{N^{2}})dr^{2}+r^{2}(d\theta^{2}+\sin^{2}\theta
d\phi^2).
\end{equation}
Comparing with the ADM metric, we find that $N(t_{PG})=1$ and if
 \be\label{nec}
g=N^{2},\ee we reach \eqref{ADMmetric}.  So the solutions
\eqref{Lu'sSolution02} of paper \cite{Lu2009} can not preserve the
``projectability condition" after the coordinate transformation. And
it seems that the solution  \eqref{Park's solution} could preserve
the ``projectability condition" after the coordinate transformation.
However note that \eqref{nec} is only a necessary condition but not
a sufficient condition. Actually from the study below, we will see
that  \eqref{Park's solution} could not satisfy the ``projectability
condition" neither.

We now seek the static, spherically symmetric solutions  with the
metric ansatz
\begin{equation}\label{ansatz-sin}
ds^2=-N(t)^{2}dt^2+\frac{1}{f(r)}(dr+N^{r}dt)(dr+N^{r}dt)+r^{2}(d\theta^{2}+\sin^{2}\theta
d\phi^2).
\end{equation}
By the coordinate transformation
$dt=dt_{s}+\frac{N_{r}}{N^{2}-fN_{r}^{2}}dr$, we can transform the
metric ansatz to the Schwarzschild coordinates type,
\begin{equation}\label{ansatz-Sch}
ds^{2}=-(N^{2}-fN_{r}^{2})dt_{S}^{2}+\frac{N^{2}}{f(N^{2}-fN_{r}^{2})}dr^{2}+r^{2}(d\theta^{2}+\sin^{2}\theta
d\phi^2).
\end{equation}

Substituting the metric ansatz \eqref{ansatz-sin} into the
Lagrangian \eqref{S_g:General_model01}, up to an overall scaling
constant, we get
\begin{eqnarray}\label{L_KV}
\nn
\mathcal{L}_{K}=&&\frac{1}{\sqrt{f}}\frac{1}{N(t)}\left\{(1-\lambda)r^{2}f^{2}\left(N^{'}_{r}+N_{r}\frac{f^{'}}{2f}\right)^{2}+2(1-2\lambda)f^{2}N_{r}^{2}\right.\\
\nn
&&\left.-4\lambda rf^{2}N_{r}\left(N^{'}_{r}+N_{r}\frac{f^{'}}{2f}\right) \right\},\\
\nn
\mathcal{L}_{V}=&&\frac{1}{\sqrt{f}}N(t)r^{2}\left\{-g_{0}\zeta^{6}+\zeta^{4}\left[\frac{2(1-f)}{r^{2}}-\frac{2f^{'}}{r}\right]+g_{2}\zeta^{2}\left[\frac{2(1-f)}{r^{2}}-\frac{2f^{'}}{r}\right]^{2}\right.\\
\nn
&&\left. +g_{3}\zeta^{2}\left[\frac{f^{'2}}{r^{2}}+\frac{2}{r^{4}}(1-f-\frac{r}{2}f^{'})^{2}\right]+g_{4}\left[\frac{2(1-f)}{r^{2}}-\frac{2f^{'}}{r}\right]^{3}\right.\\
\nn
&&\left.+g_{5}\left[\frac{2(1-f)}{r^{2}}-\frac{2f^{'}}{r}\right]\left[\frac{f^{'2}}{r^{2}}+\frac{2}{r^{4}}(1-f-\frac{r}{2}f^{'})^{2}\right]\right.\\
\nn
&&\left.+g_{6}\left[-\frac{f^{'3}}{r^{3}}+\frac{2}{r^{6}}(1-f-\frac{rf^{'}}{2})^{3}\right]+g_{7}\left[\frac{2(1-f)}{r^{2}}-\frac{2f^{'}}{r}\right]\frac{\sqrt{f}}{r^{2}}\partial_{r}\left\{\frac{1}{\sqrt{f}}r^{2}f\partial_{r}\left[\frac{2(1-f)}{r^{2}}-\frac{2f^{'}}{r}\right]\right\}\right.\\
&&\left.+g_{8}\left[f^{3}\left(\frac{f^{'}}{r^{2}f}-\frac{f^{''}}{rf}\right)^{2}+\frac{2f}{r^{4}}\left(\frac{f^{'}}{2}+\frac{rf^{''}}{2}+\frac{2(1-f)}{r}\right)^{2}\right]\right\}.
\end{eqnarray}
Here $N_{r}=N^{r}/f$ and $'$ means the derivative with respect to
$r$. The full Lagrangian is
$\mathcal{L}=\mathcal{L}_{K}+\mathcal{L}_{V}$. By varying the
action with respect to the functions $N_{r}$ , $f$ and $N(t)$, we
obtain three equations of motions,
\begin{eqnarray}\label{ByN_r}
\nn
0&=&\sqrt{f}\left\{\partial_{r}\frac{\partial\mathcal{L}}{\partial
N_{r}^{'}}-\frac{\partial\mathcal{L}}{\partial
N_{r}}\right\}\\
&=&2(1-\lambda)r^{2}f^{2}\frac{1}{N(t)}\left\{N_{r}^{''}+\frac{f^{''}}{2f}N_{r}+\frac{3}{2}\frac{f^{'}}{f}N_{r}^{'}+2\frac{N_{r}^{'}}{r}+\frac{1-2\lambda}{1-\lambda}\frac{f^{'}}{f}\frac{N_{r}}{r}-2\frac{N_{r}}{r^{2}}\right\},\\
\label{By_f} \nn
0&=&\sqrt{f}\left\{\partial_{r}\frac{\partial\mathcal{L}}{\partial
f^{'}}-\frac{\partial\mathcal{L}}{\partial
f}-\partial_{r}\partial_{r}\frac{\partial\mathcal{L}}{\partial f^{''}}\right\}\\
\nn&=&\sqrt{f}\left\{\partial_{r}\frac{\partial\mathcal{L}_{V}}{\partial
f^{'}}-\frac{\partial\mathcal{L}_{V}}{\partial
f}-\partial_{r}\partial_{r}\frac{\partial\mathcal{L}_{V}}{\partial
f^{''}}\right\}-\frac{f^{'}}{2f}\frac{1}{N(t
)}\left\{(1-\lambda)r^{2}fN_{r}\left(N_{r}^{'}+N_{r}\frac{f^{'}}{2f}\right)-2\lambda
rfN_{r}^{2}\right\}\\
\nn&&\;+\frac{1}{N(t)}\left\{(1-\lambda)r^{2}fN_{r}N_{r}^{''}+\frac{1}{2}(1-\lambda)r^{2}f^{''}N_{r}^{2}-(1-\lambda)r^{2}fN_{r}^{'2}+(1-\lambda)r^{2}f^{'}N_{r}N_{r}^{'}\right.\\
&&\;+\left.2(1+\lambda)rfN_{r}N_{r}^{'}+(1-\lambda)rf^{'}N_{r}^{2}+(6\lambda-4)fN_{r}^{2}\right\}+\frac{1}{2\sqrt{f}}\mathcal{L}_{K},\\
\label{HamiltonianConstraint01}
0&=&\int_{0}^{\infty}drr^{2}\frac{1}{N(t)}\left(-\mathcal{L}_{K}+\mathcal{L}_{V}\right).
\end{eqnarray}
The third equation \eqref{HamiltonianConstraint01} is a spatially
integrated Hamiltonian constraint because of the ``projectability
condition'' on the lapse function $N(t)$. We find that for all
$\lambda$, $N_{r}=0$ is the solution of the equation
\eqref{ByN_r}. In this case, the equations
\eqref{By_f},\eqref{HamiltonianConstraint01} are the equations
depending on the form of the potential. We can make ansatz
$f(r)=1+yr^{2}$, where $y$ is a constant to be determined. Then we
have two cubic equations of $y$
\begin{eqnarray}
g_{0}\zeta^{6}+2\zeta^{4}y+4(3g_{2}+g_{3})\zeta^{2}y^{2}-24(9g_{4}+3g_{5}+g_{6})y^{3}&=&0,\label{eq-y'sequation}\\
g_{0}\zeta^{6}+6\zeta^{4}y-12(3g_{2}+g_{3})\zeta^{2}y^{2}+24(9g_{4}+3g_{5}+g_{6})y^{3}&=&0.\label{eq-HM'Nr=0}
\end{eqnarray}
Here the equation \eqref{eq-HM'Nr=0} is from the non-local
Hamiltonian constraint.

 For the solution $f=1+yr^2$, the metric now has the form
\begin{equation}\label{maxsymm}
ds^{2}=-dt^{2}+\frac{dr^{2}}{1+yr^2}+r^{2}(d\theta^{2}+\sin^{2}\theta
d\phi^{2}).
\end{equation}
Such a metric describes a spacetime of topology $\mathbb{R}\times
\mathbb{M}_3$, where $\mathbb{M}_3$ is a three-dimensional maximally
symmetric space, could be a flat space, a sphere or a hyperboloid.
If $y=0$, this is just the flat spacetime. If $y<0$, the spacetime
is $\mathbb{R}\times \mathbb{S}^3$, where $\mathbb{R}$ is the time
direction, $\mathbb{S}^3$ is the three-sphere. If $y>0$, the
spacetime is $\mathbb{R}\times \mathbb{H}^3$, where $\mathbb{H}^3$
is the three-dimensional hyperboloid with negative constant
curvature. In fact, if one considers the time-dependent solution,
then the latter two solutions are very similar to closed and open
universe with a constant scale factor.

For a general potential, there is no solution to
\eqref{eq-y'sequation} and \eqref{eq-HM'Nr=0}. When
$\zeta=1,\;g_{0}=2\Lambda,\;g_{2}=g_{3}=g_{4}=g_{5}=g_{6}=g_{7}=g_{8}=0$,
it recovers Einstein's general relativity. The only possible
solution requires $g_0=0$ and $y=0$, which corresponds to a flat
spacetime. Actually, when the cosmological constant is vanishing,
the flat Minkowski spacetime corresponding to $y=0$ is always a
solution.

For the original \hl~ gravity with the action \eqref{S_g:origin},
the equations \eqref{eq-y'sequation},\eqref{eq-HM'Nr=0} become
\begin{eqnarray}
&&y^{2}-2\Lambda_{W}y-3\Lambda_{W}^{2}=0,\\
&&y^{2}+2\Lambda_{W}y+\Lambda_{W}^{2}=0.
\end{eqnarray}
The solution is $y=-\Lambda_{W}$. In this case, the curvature of
maximally symmetric space is determined by the cosmological
constant of the theory.

For the general action of modified \hl~gravity, the existence of
the solution depends on the form of the potential. It is easy to
see that the equations \eqref{eq-y'sequation},\eqref{eq-HM'Nr=0}
could be reduced to two equations both quadratic in $y$. It is
straightforward to find the condition under which there exist a
solution.

In the IR region,  the modified \hl~ gravity recovers the Einstein's
general relativity except the higher derivative terms on the spatial
metric. When $\lambda=1$, the equation \eqref{ByN_r} becomes
\begin{equation}
\frac{f^{'}}{f}\frac{N_{r}}{r}=0.
\end{equation}
Its solutions are $N_{r}=0$ or $f=\mbox{constant}$. The solution
$N_{r}=0$ has been discussed above. When $f$ is a constant, the
equations \eqref{By_f},\eqref{HamiltonianConstraint01} become
\begin{eqnarray}\label{eq-FromByf}
\nn
0=(N_{r}^{2})'+\frac{N_{r}^{2}}{r}+\frac{N(t)^{2}}{2f^{2}}&&\left\{-g_{0}\zeta^{6}r+\frac{2\zeta^{4}(1-f)}{r}
+\frac{2\zeta^{2}(1-f)}{r^{3}}\left[2g_{2}(1+7f)+g_{3}(1+5f)\right]\right.\\
\nn &&
\left.+\frac{2(1-f)^{2}}{r^{5}}\left[4g_{4}(1+23f)+2g_{5}(1+17f)+g_{6}(1+14f)\right]\right.\\
&&\left.+\frac{8f(1-f)}{r^{5}}\left[2g_{7}(1+7f)+g_{8}(1-4f)\right]\right\},
\end{eqnarray}
\begin{eqnarray}\label{eq-FromHL}
\nn 0&=&\int_{0}^{\infty}drr^{3}\left\{(N_{r}^{2})'+\frac{N_{r}^{2}}{r}+\frac{N(t)^{2}}{2f^{2}}\left[-g_{0}\zeta^{6}r
+\frac{2\zeta^{4}(1-f)}{r}+\frac{2\zeta^{2}(1-f)^{2}}{r^{3}}\left(2g_{2}+4g_{3}\right)\right.\right.\\
&&\left.\left.+\frac{2(1-f)^{3}}{r^{5}}\left(4g_{4}+2g_{5}+g_{6}\right)+\frac{8f(1-f)^{2}}{r^{5}}\left(g_{7}+g_{8}\right)\right]\right\}.
\end{eqnarray}
It is not hard to find that just when $f=1$ the two equations have
the same solutions of $N_{r}$. In other words, $f$ is constrainted
to be $1$. In this case, the solutions are just
\begin{equation}\label{By_f_f=1}
N_{r}=\pm \;N(t)\sqrt{\frac{g_{0}\zeta^{6}}{6}r^{2}+\frac{M}{r}},
\end{equation}
where $M$ is an integration constant. For $N_{r}$  is just the
function of $r$, $N(t)$ must be a constant. We could use the
freedom of gauge transformation to set $N(t)=1$. If let
$g_{0}\zeta^{6}=3\Lambda_{W}$, the solution \eqref{By_f_f=1}
corresponds to a dS-Schwarzschild spacetime written in
Painlev\'e-Gullstrand type coordinates. The solution is just
determined by the kinetic term and the cosmological constant in
the potential. In other words, at IR, the static spherically
symmetric solutions of the modified \hl~gravity are the same as
the ones in the Einstein's general relativity. If the theory has a
nonvanishing cosmological constant, the solution is the
Schwarzschild solution in dS spacetime. If the theory has no
cosmological constant, the solution is just the Schwarzschild
solution.

In the UV region when $\lambda\neq 1$, similar to the discussion
in the IR region, the equations \eqref{ByN_r}, \eqref{By_f} and
\eqref{HamiltonianConstraint01} have solutions just when $f=1$. In
this case, they become
\begin{eqnarray}
0&=&N_{r}^{''}+2\frac{N_{r}^{'}}{r}-2\frac{N_{r}}{r^{2}},\label{By_Nr_f1}\\
0&=& (1-\lambda)r^{2}N_{r}^{'2}-4\lambda
rN_{r}N_{r}^{'}+2(1-2\lambda)N_{r}^{2}+g_{0}N(t)^{2}\zeta^{6}r^{2},\label{By_f_f1}\\
0&=&\int_{0}^{\infty}drr^{2}\left\{(1-\lambda)r^{2}N_{r}^{'2}-4\lambda
rN_{r}N_{r}^{'}+2(1-2\lambda)N_{r}^{2}+g_{0}N(t)^{2}\zeta^{6}r^{2}\right\}.\label{eq-byHL02}
\end{eqnarray}
 They have  solutions as
 \be
 N_{r}=\pm\;
 N(t)\sqrt{\frac{g_{0}\zeta^{6}}{3(3\lambda-1)}}r.\label{solutions-lambda-eq1}
 \ee
We could also use the freedom of gauge transformation to set
$N(t)=1$. These solutions actually describe the same de-Sitter
space-time. One easy way to see this point is to change inversely
into the Schwarzschild coordinates.

One subtle issue happens when the cosmological constant $\L_W$ is
negative. In this case, $N_r$ becomes imaginary in
\eqref{solutions-lambda-eq1}. This is not physical anymore.
However, after being transformed into Schwartzschild coordinates,
the metric describes the anti-de-Sitter spacetime. Similarly the
solution \eqref{By_f_f=1} becomes imaginary at asymptotic region
if $\L_W$ is negative, but it may describe a AdS-Sch. spacetime in
the Schwarzschild coordinates. Since in \hl-like gravity, to
respect the projectability condition, the static spherically
symmetric solution should take the form of \eqref{ansatz-sin}, the
solutions with negative $\L_W$ are not acceptable. It would be
interesting to see if the AdS and AdS-Sch. spacetime could be
rewritten into  a form respecting projectability
condition\footnote{In \cite{Lu2009}, it has been pointed out that
the dS-Sch. solution could be rewritten in terms of the
Painlev\'e-Gullstrand  coordinates to respect the projectability
condition. We are also grateful to H.Lu for the discussion on the
pathology of negative $\L_W$.}.

 After some tedious calculation, it is straightforward  to check that the
solutions \eqref{maxsymm},\eqref{By_f_f=1}, and
\eqref{solutions-lambda-eq1} satisfy all the equations of $\delta
S/\delta N(t)=0$, $\delta S/\delta N_{i}=0$ and $\delta S/\delta
g_{ij}=0$. Obviously they are all the solutions of \horava\
gravity in the IR region($\lambda=1$). So the new solutions found
in \cite{Lu2009,park2009} could not satisfy the ``projectability
condition", even though they satisfy the necessary condition
\eqref{nec}. Our result also indicates that in \hl-like gravity
theory with the projectability condition, there is no novel
correction in solar system test.

It is also interesting to study the topological black hole in
\hl~like gravity. It has been discussed in \cite{RongCai-2009}
without taking into account of the ``projectability condition". The
static spherically symmetric metric ansatz  of a topological
spacetime may be written as
\begin{equation}\label{topologicalansatz}
ds^2=-dt^2+\frac{1}{f(r)}(dr+N^{r}dt)(dr+N^{r}dt)+r^{2}d\Omega_{k}^{2}
\end{equation}
Here we have set $N(t)=1$ and $d\Omega_{k}^{2}$ denotes the line
element for an 2-dimensional Einstein space with constant scalar
curvature $2k$. Without loss of generality, one may take $k=0,\pm1$
respectively. Substituting the metric ansatz
\eqref{topologicalansatz} into the Lagrangian
\eqref{S_g:General_model01}, up to an overall scaling constant, we
get
\begin{eqnarray}\label{L_KV-topo}
\nn
\mathcal{L}_{K}=&&\frac{1}{\sqrt{f}}\left\{(1-\lambda)r^{2}f^{2}\left(N^{'}_{r}+N_{r}\frac{f^{'}}{2f}\right)^{2}+2(1-2\lambda)f^{2}N_{r}^{2}\right.\\
\nn
&&\left.-4\lambda rf^{2}N_{r}\left(N^{'}_{r}+N_{r}\frac{f^{'}}{2f}\right) \right\},\\
\nn
\mathcal{L}_{V}=&&\frac{1}{\sqrt{f}}r^{2}\left\{-g_{0}\zeta^{6}+\zeta^{4}\left[\frac{2(k-f)}{r^{2}}-\frac{2f^{'}}{r}\right]+g_{2}\zeta^{2}\left[\frac{2(k-f)}{r^{2}}-\frac{2f^{'}}{r}\right]^{2}\right.\\
\nn
&&\left. +g_{3}\zeta^{2}\left[\frac{f^{'2}}{r^{2}}+\frac{2}{r^{4}}(k-f-\frac{r}{2}f^{'})^{2}\right]+g_{4}\left[\frac{2(k-f)}{r^{2}}-\frac{2f^{'}}{r}\right]^{3}\right.\\
\nn
&&\left.+g_{5}\left[\frac{2(k-f)}{r^{2}}-\frac{2f^{'}}{r}\right]\left[\frac{f^{'2}}{r^{2}}+\frac{2}{r^{4}}(k-f-\frac{r}{2}f^{'})^{2}\right]\right.\\
\nn
&&\left.+g_{6}\left[\frac{f^{'3}}{r^{3}}+\frac{2}{r^{6}}(k-f-\frac{rf^{'}}{2})^{3}\right]+g_{7}\left[\frac{2(k-f)}{r^{2}}-\frac{2f^{'}}{r}\right]\frac{\sqrt{f}}{r^{2}}\partial_{r}\left\{\frac{1}{\sqrt{f}}r^{2}f\partial_{r}\left[\frac{2(k-f)}{r^{2}}-\frac{2f^{'}}{r}\right]\right\}\right.\\
&&\left.+g_{8}\left[f^{3}\left(\frac{f^{'}}{r^{2}f}-\frac{f^{''}}{rf}\right)^{2}+\frac{2f}{r^{4}}\left(\frac{f^{'}}{2}+\frac{rf^{''}}{2}+\frac{2(k-f)}{r}\right)^{2}\right]\right\}.
\end{eqnarray}
Here $N_{r}=N^{r}/f$ and $'$ means the derivative with respect to
$r$. The full Lagrangian is
$\mathcal{L}=\mathcal{L}_{K}+\mathcal{L}_{V}$. The $k=1$ case has
been discussed above. Comparing with \eqref{L_KV}, we find that the
kinetic term is exactly the same, and the difference in the
potential term coming from the factor $(k-f)$ in \eqref{L_KV-topo}
and $(1-f)$ in \eqref{L_KV}. By varying the action with respect to
the functions $N_{r}$, $f$ and $N(t)$, we could get three equations
of motions which are quite similar to \eqref{ByN_r},\eqref{By_f} and
\eqref{HamiltonianConstraint01}, with $(1-f)$ being replaced with
$(k-f)$. Therefore the solutions are quite similar to the ones when
$k=1$.

The case $k=1$ has been discussed above. In  the case $k=-1$, for
the solution with $f$ being a constant, $f$ must be set to $-1$.
At IR, $\lambda=1$,  $N_{r}=\pm
\;\sqrt{\frac{g_{0}\zeta^{6}}{6}r^{2}+\frac{M^{\star}}{r}}$, where
$M^{\star}$ is an integration constant. They correspond to an
(dS-)Schwarzschild type's topological black hole written in
Painlev\'e-Gullstrand type coordinates. When $\lambda\neq 1$,
$N_{r}=\pm \sqrt{\frac{g_{0}\zeta^{6}}{3(3\lambda-1)}}r$. These
solutions actually describe the de-Sitter space-time or Minkowski
spacetime with topological twist. In the case $k=0$, because $f$
can't be zero, we only have the solution ``$N_{r}=0,\,f=yr^{2}$''
in which $y$ satisfy the equation
\eqref{eq-y'sequation},\eqref{eq-HM'Nr=0}. In any case, these
solutions are different from the ones studied in
\cite{RongCai-2009}.

\section*{Acknowledgments}
The work was partially supported by NSFC Grant No.10535060,
10775002, 10975005 and RFDP.


\begin{thebibliography}{99}

\bibitem{Horava:2008ih}
  P.~Horava,
  ``Membranes at Quantum Criticality,''
  JHEP {\bf 0903}, 020 (2009)
  [arXiv:0812.4287 [hep-th]].

\bibitem{Horava:2009uw}
  P.~Horava,
  ``Quantum Gravity at a Lifshitz Point,''
  Phys.\ Rev.\  D {\bf 79}, 084008 (2009)
  [arXiv:0901.3775 [hep-th]].

\bibitem{Lifshitz}E.M. Lifshitz, ``On the Theory of Second-Order
Phase Transitions I \& II", Zh. Eksp. Teor. Fiz {\bf 11} (1941)255
\& 269.

\bibitem{Chen:2009ka}
  B.~Chen and Q.~G.~Huang,
  ``Field Theory at a Lifshitz Point,''
  arXiv:0904.4565 [hep-th].


\bibitem{Cai:2009dx}
  R.~G.~Cai, B.~Hu and H.~B.~Zhang,
  ``Dynamical Scalar Degree of Freedom in Horava-Lifshitz Gravity,''
  Phys.\ Rev.\  D {\bf 80}, 041501 (2009)
  [arXiv:0905.0255 [hep-th]].

\bibitem{Chen:2009jr}
  B.~Chen, S.~Pi and J.~Z.~Tang,
  ``Scale Invariant Power Spectrum in Ho\v{r}ava-Lifshitz Cosmology without
  Matter,'' JCAP {\bf 08} (2009)007,
  arXiv:0905.2300 [hep-th].

\bibitem{Li:2009bg}
  M.~Li and Y.~Pang,
  ``A Trouble with Ho\v{r}ava-Lifshitz Gravity,''
  arXiv:0905.2751 [hep-th].


\bibitem{Lu2009}H.~Lu, J.~Mei and C.~N.~Pope,
  ``Solutions to Horava Gravity,''
  arXiv:0904.1595 [hep-th].

\bibitem{Nastase2009} Horatiu Nastase,
  ``On IR solutions in Horava gravity theories,''
  arXiv:0904.3604 [hep-th]

\bibitem{Kehagias:2009is}
  A.~Kehagias and K.~Sfetsos,
  ``The black hole and FRW geometries of non-relativistic gravity,''
  Phys.\ Lett.\  B {\bf 678}, 123 (2009)
  [arXiv:0905.0477 [hep-th]].


\bibitem{AhmadGhodsi2009}   Ahmad Ghodsi ,
  ``Toroidal solutions in Horava Gravity,''
  arXiv:0905.0836 [hep-th].


\bibitem{park2009}   Mu-in Park,
  ``The Black Hole and Cosmological Solutions in IR modified Horava Gravity,''
  arXiv:0905.4480 [hep-th]

\bibitem{Colgain:2009fe}
  E.~O.~Colgain and H.~Yavartanoo,
  ``Dyonic solution of Horava-Lifshitz Gravity,''
  JHEP {\bf 0908}, 021 (2009)
  [arXiv:0904.4357 [hep-th]].


  \bibitem{Solar}T.~Harko, Z.~Kovacs and F.~S.~N.~Lobo,
  ``Testing Ho\v{r}ava-Lifshitz gravity using thin accretion disk properties,''
  Phys.\ Rev.\  D {\bf 80}, 044021 (2009)
  [arXiv:0907.1449 [gr-qc]].\\
 T.~Harko, Z.~Kovacs and F.~S.~N.~Lobo,
  ``Solar system tests of Ho\v{r}ava-Lifshitz gravity,''
  arXiv:0908.2874 [gr-qc].\\
  L.~Iorio and M.~L.~Ruggiero,
  ``Horava-Lifshitz gravity and Solar System orbital motions,''
  arXiv:0909.2562 [gr-qc].



\bibitem{Visser_2009} Thomas P. Sotiriou, Matt Visser, Silke Weinfurtner,
 ``Phenomenologically viable Lorentz-violating quantum gravity",
 Phys.Rev.Lett.{\bf 102}:251601,2009,
arXiv:0904.4464 [hep-th];``Quantum gravity without Lorentz
invariance,''
 arXiv:0905.2798 [hep-th].



\bibitem{Painleve} Painlev\'e P. La m\'ecanique classique el la theorie de la
relativit\'e(Classical mechanics of the theory of relativity).C. R.
Acad. Sci. (Paris), 173 (1921), 677¨C680.


\bibitem{Gullstrand} Gullstrand A. Allegemeine l$\ddot{o}$sung des statischen eink$\ddot{o}$rper-problems in
der einsteinshen gravitations theorie (General solution for static
onebody problems in Einstein¡¯s theory of gravity).\emph{Arkiv. Mat.
Astron. Fys.},16(8) (1922), 1-15.

\bibitem{Lematre} Lema$\hat{i}$tre G. L'nivers en expansion (The universe in
expansion).\emph{Ann. Soc. Sci. (Bruxelles)},A53 (1933), 51-85.

\bibitem{Hawking} Hawking S W and Israel S W (editors).\emph{Three
hundred years of gravitation}. Cambridge University Press, England
(1987). See especially the discussion on page 234.

\bibitem{RongCai-2009} Rong-Gen Cai, Li-Ming Cao, Nobuyoshi Ohta,
``Topological Black Holes in Horava-Lifshitz Gravity,''
 Phys.\ Rev.\  D {\bf 80}, 024003 (2009)
  [arXiv:0904.3670 [hep-th]]

\bibitem{A.A.Kocharyan}  A.A.Kocharyan,
``Is nonrelativistic gravity possible?''
 Phys.\ Rev.\  D {\bf 80}, 024026 (2009)
  [arXiv:0905.4204 [hep-th]]


\end{thebibliography}
\end{document}